# Fully compensated ferrimagnetic triferroics and multistate transport in hidden-phase wurtzite MnSe monolayer

Zhuang Ma[1], Hongfei Liang[2], Po Ma[2], Guangqian Ding[3], Xuehao Wu[4], Sikander Azam[5], Guoying Gao[2], and Long Zhang[2,*]

[1]*School of Physics and Telecommunication Engineering*, *Zhoukou Normal University*, *Zhoukou 466001*, *People's Republic of China*

[2]*School of Physics and Wuhan National High Magnetic Field Center*, *Huazhong University of Science and Technology*, *Wuhan 430074*, *People's Republic of China*

[3]*School of Electronic Science and Engineering, Chongqing University of Posts and Telecommunications, Chongqing 400065, People's Republic of China*

[4]*Department of Physics, Columbia University, New York, NY 10027, USA*

[5]*New Technologies—Research Centre, University of West Bohemia, Univerzitni 8, Pilsen 306 14, Czech Republic*

*Contact author: longzhangphysics@foxmail.com

Fully compensated ferrimagnets (fFIMs) have attracted interest due to their compensated moments and nonrelativistic spin splitting across the Brillouin zone. Known fFIMs, however, are mostly restricted to complex three-dimensional (3D) systems or require external fields in two-dimensional (2D) heterostructures, leaving intrinsic fFIM monolayers unexplored. We identify a hidden-phase MnSe monolayer, derived from the (001) planes of wurtzite, as an intrinsic fFIM featuring inequivalent sublattices not linked by any symmetry. It is a unipolar magnetic semiconductor (UMS) with perpendicular magnetic anisotropy (528.60 μeV per unit cell) and simultaneously exhibits ferroelectricity (polarization $4.63 \times 10^{-10}$ C/m) and ferroelasticity (signal 61%), with barriers of $7.6 \times 10^{-3}$ and 0.10 eV/f.u., respectively, establishing a single-phase triferroic system. The ground fFIM UMS characteristics are robust against strain up to ±3%. The $In_2Se_3$/MnSe heterostructure enables nonvolatile electrical control between semiconducting and metallic states. Constructed tunnel junctions exhibit giant tunneling magnetoresistance ($2.98 \times 10^5$%), electroresistance ($6.97 \times 10^{14}$%), elastoresistance ($7.95 \times 10^4$%), and near-perfect spin filtering (~100%). Collectively, this spontaneous 2D fFIM with coexisting triferroic orders provides a promising platform for ultrahigh-density, low-power, and miniaturized memory devices.

## I. INTRODUCTION

Ferromagnets (FMs) exhibit net magnetization and spin-split bands, whereas antiferromagnets (AFMs) possess compensated moments with Kramers degeneracy. These conventional phases are limited by stray fields in FMs and weak spin responses in AFMs. Recently, altermagnets (AMs) and fully compensated ferrimagnets (fFIMs) have emerged as zero-net-moment, spin-polarized collinear phases [1-5]. AMs break $\tau T/PT$ ($T$, $P$, and $\tau$ denote time-reversal, inversion, and translation, respectively), with opposite-spin sublattices connected by rotation ($C$) or mirror ($M$) symmetry [6]. In fFIMs, however, the sublattices are not symmetry-related; compensation arises from electron filling, yielding *s*-wave-like global spin splitting across the entire Brillouin zone (BZ) [Fig. 1(a)]. This enables fully spin-polarized currents [7] while retaining the advantages of compensated order, holding promise for high-density device integration and information storage. Yet, fFIMs have been realized primarily in complex three-dimensional (3D) systems (e.g., double perovskite oxides, Heusler compounds, and organic materials) [8-12], or through external perturbations (e.g., electric field, alloying, stacking, and adsorption) in 2D van der Waals (vdW) materials [13-16]; intrinsic 2D fFIMs remain largely uncharted.

Combining $T$ symmetry breaking (magnetism) with broken structural symmetry can induce ferroelectricity and ferroelasticity. Reported magnetic multiferroics, such as $CrI_2$ [17], $FeO_2H$ [18], and $CrCoS_4$ [19], enable multifield-controllable devices. Magnetic alignment (parallel or antiparallel) and ferroelectric polarization switching yield tunneling magnetoresistance (TMR) and tunneling electroresistance (TER) [20], respectively. Similarly, ferroelastic transition (structure deformation) can induce tunneling elastoresistance (TAR). Spin-filtering effect can enhance the read/write fidelity via pure spin currents. Despite these prospects, multiferroicity, and particularly triferroicity, in fFIMs is scarcely reported, with their triferroic transport behavior even less explored.

Non-vdW flakes, which possess superior thermal and electrical conductivities favorable for high integration, can be prepared via liquid exfoliation [21,22] or chemical vapor deposition (CVD) [23]. Recently, wurtzite-type MnSe bulk and its ultrathin nanosheets have been synthesized via CVD [24]. Hidden ferroelectric phases in wurtzite (001) planes [25] have been demonstrated, and similar 2D structures can be constructed from triangular and tetrahedral building blocks [26]. Here, we select the magnetic MnSe monolayer as a representative structure to explore potential intrinsic multiferroics and transport performances [Fig. 1(b)].

In this work, we systematically investigate the structural stability, electronic structure, ferroic properties, magnetic anisotropy, and multistate transport of the hidden-phase wurtzite MnSe monolayer. It possesses thermal, dynamic, and mechanical stabilities. This monolayer is a fFIM and exhibits intrinsic ferroelectricity and ferroelasticity with moderate transition barriers. Uniaxial and biaxial strains can effectively tune the band gap, spin splitting, and magnetic anisotropy energy (MAE), whereas the fully compensated magnetic moments remain robust throughout the strain range. The proposed triferroic tunnel junctions (TFTJs) exhibit TMR of $2.28 \times 10^{2}$%-$2.98 \times 10^{5}$%, TER of $2.93 \times 10^{6}$%-$6.97 \times 10^{14}$%, and TAR of $4.63 \times 10^{2}$%-$7.95 \times 10^{4}$%, along with ~100% spin filtering efficiency. These results motivate developing advanced triferroic applications.

## II. COMPUTATIONAL DETAILS

The density functional theory (DFT) calculations [27] for structure optimization, total energy, and electronic structure were performed using the Vienna *ab initio* Simulation Package (VASP) codes [28,29]. The exchange-correlation effects were described by the Perdew-Burke-Ernzerhof (PBE) functional within the generalized gradient approximation (GGA) [30]. Following previous reports [31,32], an effective Hubbard correction $U_{\mathrm{eff}}$ = 3.0 eV was applied to Mn-3*d* orbitals. The electron-electron and electron-ion interactions were described via the projector augmented-wave potentials [33]. A plane-wave cutoff energy of 500 eV was adopted, and the total energy and force convergence criteria were chosen as $10^{-7}$ eV and 0.001 eV/Å, respectively.

The BZ was sampled with Γ-centered *k*-meshes of 8 × 5 × 1 for the monolayer, 6 × 4 × 1 for the heterostructure, and 12 × 12 × 6 for the bulk wurtzite phase. Mn-3*d*, 4*s* and Se-4*s*, 4*p* were considered. A 20 Å vacuum layer was employed along the *z* direction to eliminate the interactions between periodic images of the 2D structures.

Phonon dispersion spectra were calculated using the PHONOPY code based on density functional perturbation theory (DFPT) with a 4 × 4 × 1 supercell [34]. *Ab initio* molecular dynamics (AIMD) simulation was performed on a 4 × 4 × 1 supercell at 300 K using a time step of 3 fs. The MAE was defined as $E_{MAE} = E_{SOC}^{\parallel} - E_{SOC}^{\perp}$, where $E_{\parallel}$ and $E_{\perp}$ denote the total energies for magnetization axes lying in the plane and perpendicular to the plane, respectively. The solid-state nudged elastic band (SSNEB) method [35] was employed to calculate the transition energy barrier, which directly maps the minimum energy pathways associated with lattice deformation observed in experiments. To evaluate electric polarization, the Berry phase method [36] was used to calculate the spontaneous polarization of the MnSe monolayer. Visualized crystal structures and the IDPP-based ferroelastic switching pathways were obtained using the ATOMKIT package [37,38].

The transport properties of the $In_2Se_3$/MnSe heterostructures were calculated using the QuantumATK code [39] based on DFT combined with the nonequilibrium Green's function (NEGF) approach. The cutoff energy was set to be 150 hartree, and the *k*-point grids of 1 × 10 × 100 and 100 × 100 were selected for self-consistent and transmission calculations, respectively. The spin filtering efficiency is defined as

$$\eta = \left| \frac{T_{\uparrow} - T_{\downarrow}}{T_{\uparrow} + T_{\downarrow}} \right| \times 100\% \tag{1}$$

where $T_{\uparrow}$ and $T_{\downarrow}$ denote the spin-up and spin-down transmission coefficients, respectively. TMR is given by

$$\mathrm{TMR} = \frac{T_{\mathrm{M_P}} - T_{\mathrm{M_{AP}}}}{T_{\mathrm{M_{AP}}}} \times 100\% \tag{2}$$

with $T_{\mathrm{M_P}}$ and $T_{\mathrm{M_{AP}}}$ standing for transmission coefficients for parallel and antiparallel magnetization alignments, respectively. Similarly, TER and TAR are obtained as

$$\text{TER}=\frac{|T_{\mathrm{P}_\uparrow}-T_{\mathrm{P}_\downarrow}|}{\min(T_{\mathrm{P}_\uparrow},T_{\mathrm{P}_\downarrow})}\times 100\%\,,\qquad \text{TAR}=\frac{|T_a-T_b|}{\min(T_a,T_b)}\times 100\% \tag{3}$$

in which $T_{\mathrm{P}_\uparrow}$ and $T_{\mathrm{P}_\downarrow}$ are the transmission coefficients with opposite ferroelectric polarization directions, respectively. $T_a$ and $T_b$ are the transmission coefficients with perpendicular transport directions (corresponding to two ferroelastic states), respectively. Similar computational methods have been applied in our previous studies [40,41].

## III. RESULTS AND DISCUSSION

### A. Structural stabilities and magnetism

The top and side views of the monolayer MnSe are depicted in Fig. 2(a). The primitive unit cell possesses a space group of $Pmc2_1$ (No. 26) and contains four Mn and four Se atoms, with Mn in tetrahedral (4-fold) and trigonal (3-fold) Se coordination, forming an orthorhombic lattice with broken $P$ symmetry. Among four magnetic configurations (FM, AFM1–AFM3), whose configurations and spin charge densities are calculated in Fig. S1(a) and S1(b) [42], the AFM2 is lowest in energy. The energy differences between FM and AFM1/AFM2/AFM3 orderings are 0.32/0.37/0.14 eV/unit cell. Magnetic moments are predominantly localized around the Mn sites [Fig. S1(b) [42]]. In the ground AFM2 state, moments of $\pm 4.47$ $\mu_B$ reside on Mn atoms, while those of Se are negligible (~0.003 $\mu_B$). The two opposite magnetic sublattices cannot be related by any symmetry operation owing to the distinct (triangular and tetrahedral) coordination environments [Fig. S2 [42]], thus establishing the MnSe monolayer as an intrinsic fFIM.

The optimized lattice constants are $a$ = 4.19 Å and $b$ = 6.74 Å. Phonon dispersion spectra [Fig. 2(b)] show no imaginary modes, indicating that MnSe is dynamically stable. Owing to the eight atoms in the unit cell, the phonon dispersion spectra comprise 24 vibrational branches. The three acoustic modes at low frequencies derive predominantly from the heavier Se atoms. The 21 optical modes lie at higher frequencies. Medium-frequency bands involve mixed vibrations of Mn and Se atoms, whereas the highest-frequency region is dominated by Mn-derived vibrations, with Se atoms contributing to a lesser extent. The maximum frequency is 8.04 THz, which is much smaller than that of 2D $MoS_2$ (~14 THz) [43], reflecting the soft Mn-Se bonding. The AIMD simulations at 300 K are performed to evaluate the thermal stability. After a 10 ps simulation, the final crystal snapshot [Fig. 2(c)] preserves structural integrity without bond breaking or lattice collapse, and the total energy fluctuates slightly. This confirms thermal stability. The 2D BZ of MnSe monolayer is shown in Fig. 2(d). The elastic constants ($C_{11}$ = 29.66 N/m, $C_{22}$ = 26.71 N/m, $C_{12}$ = 4.93 N/m, and $C_{66}$ = 7.27 N/m) satisfy the Born-Huang criteria ($C_{11}$, $C_{22}$, $C_{66} > 0$; $C_{11} + C_{22} - 2C_{12} > 0$) [44], indicating the mechanical stability. The in-plane Young's modulus $Y(\theta)$ and Poisson's ratio $v(\theta)$ are given by

$$Y(\theta) = \frac{C_{11}C_{22} - C_{12}^2}{C_{11}\sin^4\theta + A\sin^2\theta\cos^2\theta + C_{22}\cos^4\theta} \tag{4}$$

$$v(\theta) = \frac{C_{12}\sin^4\theta - B\sin^2\theta\cos^2\theta + C_{12}\cos^4\theta}{C_{11}\sin^4\theta + A\sin^2\theta\cos^2\theta + C_{22}\cos^4\theta} \tag{5}$$

where $A = (C_{11}C_{22} - C_{12}^2)/C_{66} - 2C_{12}$ and $B = C_{11} + C_{22} - (C_{11}C_{22} - C_{12}^2)/C_{66}$. The orientation-dependent $Y(\theta)$ and $v(\theta)$ are obtained as $\theta$ spans a whole 360° circle [Figs. 2(e), 2(f)]. Notably, MnSe exhibits the largest (smallest) Young's modulus of 28.75 N/m (20.16 N/m) along the $x$ (47° to the $x$ axis) direction, while for Poisson's ratio, the largest (smallest) values occur along the $y$ (45° to the $x$ axis) direction. These Young's modulus values are lower than those of $MoS_2$ (124.5 N/m) [45] and BN (318 N/m) [46], and comparable to other 2D magnetic materials such as monolayer $CrI_3$ (23.85 N/m) [47] and MnSeTe (42 N/m) [48], indicating that MnSe is sufficiently flexible for ferroelastic switching under strain while preserving mechanical integrity. The anisotropic mechanical response is further revealed by the Poisson's ratios in Fig. 2(f). Specifically, MnSe's $v(\theta)$ ranges from 0.17 to 0.39, similar to values reported for most 2D materials (0-0.5) [49,50], indicating high anisotropy and rigidity, and suggesting applications in flexible electronics.

Band structures reveal anisotropic spin splitting across the entire BZ [Fig. 3(a)], unlike AMs with spin splitting limited to some high-symmetry points [2,51]. Both the conduction band minimum (CBM) at the Γ point and the valence band maximum (VBM) at the Y point near the Fermi level ($E_F$) are dominated by spin-up channel. This system is a unipolar magnetic semiconductor (UMS), which is also known as half-semiconductor (HSC) [52], with spin-up, spin-down, and total band gaps of 1.82, 2.20, and 1.82 eV, respectively. This spin polarization enables low-power spin injection, promising energy-efficient spintronic devices like spin injectors and transistors [53,54]. Specifically, the spin splitting values are 0.49 eV, 0.12 eV, and 0.20 eV at the VBM (Y), VBM (Γ), and CBM (Γ), respectively [Fig. 3(d)]. The spin texture near Y and Γ exhibits strong out-of-plane Zeeman-like polarization [55]. Projected density of states (PDOS) [Fig. 3(b)] indicates that states near the $E_F$ mainly arise from Mn-3*d* and Se-4*p* hybridization. The integrated projected density of states (IPDOS) for spin-up (Mn-3*d*-up, Se-4*p*-up) and spin-down (Mn-3*d*-down, Se-4*p*-down) shows distinct behavior across the entire energy range. Within the band gap, the total spin-up and spin-down IPDOS values are strictly equal in magnitude but opposite in sign, resulting in a net zero magnetic moment [Fig. 3(c)], supporting the fFIM nature. Notably, bulk wurtzite MnSe was predicted to be a room-temperature altermagnet [56], but its monolayer becomes a fFIM owing to quantum confinement and symmetry reduction.

MAE is crucial for stabilizing the long-range magnetic ordering in 2D systems [57]. The polar angle dependence of MAE [Fig. 3(e)] shows an out-of-plane easy axis, with energies of 528.60 (*x*-axis), 233.06 (*y*-axis) μeV per unit cell relative to the *z* axis. The "peanut"-shaped 3D anisotropic MAE map confirms the perpendicular magnetic anisotropy [Fig. 3(f)]. Orbital-resolved analysis exposes the main MAE contribution from the hybridization of Mn-3*d* orbitals [Figs. 3(g) and S3(b) [42]]. The obviously small and negative contribution is mainly dominated by the hybridization between $d_{xy}$ and $d_{x^2-y^2}$, $d_{xz}$ and $d_{yz}$, $d_{xz}$ and $d_{x^2-y^2}$ orbitals. The large positive contribution originates from the interactions between $d_{xy}$ and $d_{xz}$, $d_{yz}$ and $d_{z^2}$, $d_{yz}$ and $d_{x^2-y^2}$ orbitals for $E_x$ - $E_z$. Comparing $E_y$ with $E_z$, the hybridization between $d_{xy}$ and $d_{yz}$, $d_{xz}$ and $d_{z^2}$, $d_{xz}$ and $d_{x^2-y^2}$ all yield positive contributions to the MAE. The remaining interactions between $d_{xy}$ and $d_{x^2-y^2}$, $d_{xz}$ and $d_{yz}$ provide negative contributions to the MAE.

### B. Ferroelectricity and ferroelasticity

The MnSe monolayer ($C_{2v}$ point group) exhibits cooperative off-centering displacements of Mn atoms that break spatial symmetries (e.g., $x$, $y$, $z \rightarrow x$, $-y$, $z$), producing a spontaneous ferroelectric polarization of $4.63 \times 10^{-10}$ C/m along the $b$ axis [Fig. 4(a)] after accounting for the polarization quantum [58]. This value, which is larger than in many known 2D ferroelectrics, such as $\beta$-GeSe ($1.59 \times 10^{-10}$ C/m) [59], $FeHfSe_3$ ($1.29 \times 10^{-10}$ C/m) [60], and $(CrBr_3)_2Li$ ($0.92 \times 10^{-10}$ C/m) [61], can be attributed to the large cooperative in-plane displacements of Mn-Se pairs. As shown in the double-well profile [Fig. 4(a)], the ferroelectric $Pmc2_1$ phase is 1.03 eV/f.u. lower than the paraelectric *Pmma* phase, with a switching barrier of 0.0076 eV/f.u. (close to that of bulk, 0.0077 eV/f.u.) [Fig. 4(b)], because the 2D lattice expansion reduces bonding but enhances ionic displacement. The spin splitting is not reversed by ferroelectric switching [Figs. 3(a) and S4(a) [42], classifying the MnSe monolayer as type-I multiferroic, yet both electric dipoles and magnetic moments originate from the same Mn ions, unlike conventional type-I multiferroics where ferroelectricity and magnetism arise from different ionic species and thus exhibit weak magnetoelectric coupling. The single-phase type-I multiferroic with a common microscopic origin for electric dipoles and magnetic moments has been demonstrated [61,62]. Coexisting electric polarization and magnetic order allows multifield manipulation of charge and spin degrees of freedom, which is highly desirable for electronic and spintronic applications.

Ferroelasticity arises from two degenerate orientation variants of the $Pmc2_1$ structure, derived from a higher-symmetry paraelastic *Cmma* parent phase (in-plane $a' = b' = 5.47$ Å). The equally stable ferroelastic states can be switched from one to another under external strain or higher-order electric field [18,63]. Switching between variants involves a 90° rotation of the lattice, with a 0.10 eV/f.u. barrier [Fig. 4(c)], which is comparable to that of the 2D ferroelastic $ZrI_2$ (0.07 eV/f.u.) [64]. The ferroelastic signal intensity can be estimated by reversible ferroelastic strain, defined as $(|b|/|a|) - 1 \times 100\%$. It is calculated as 61%, indicating a strong ferroelastic signal for MnSe, comparable to those of VSSe (73%) [65] and $BP_5$ (41.4%) [66]. The spontaneous lattice strain between the initial invariant $F$ and final invariant $F''$ can be described by transformation strain matrix $\eta$, taking the paraelastic phase as a reference. Based on the Green-Lagrange strain tensor [67], the transformation strain matrix is

$$\eta = \frac{1}{2}\left[ (H_P^{-1})^T H_I^T H_I H_P^{-1} - I \right] \tag{6}$$

where the superscript -1 indicates the matrix inversion, $T$ denotes matrix transpose, $I$ is a 2 × 2 identity matrix, and $H_I$ and $H_P$ are the matrices of the ferroelastic and paraelastic states, respectively. For MnSe monolayer, the matrices $H_I$ and $H_P$ are [4.19, 0; 0, 6.74] and [5.47, 0; 0, 5.47]. The corresponding strain matrix $\eta$ can be calculated to be [-0.21, 0; 0, 0.26], suggesting that there is 21% compressive strain along the $x$ direction and 26% tensile strain along the $y$ direction. These values are comparable with those of other 2D ferroelastic materials such as stanene (21% and 36%) [68] and AlSCl (14.5% and 24.4%) [69]. The electronic band anisotropy is effectively switched upon the ferroelastic phase transition [Fig. S4(b) [42]], which is useful for anisotropic transport.

### C. Strain and heterostructure engineering

Lattice mismatch commonly occurs when 2D materials are grown on substrates [70]. The effects of experimentally feasible strain ranging from -3% (compressive) to 3% (tensile) are examined. Both the FM-AFM1 and FM-AFM2 energy differences are enhanced (reduced) by compressive (tensile) strain, while the FM-AFM3 energy difference remains nearly constant (~0.15 eV) [Figs. 5(a-c)]. This indicates that compressive strain enhances the relative stability of AFM1 and AFM2 phases against the FM phase, while tensile strain reduces their relative stability. The strain-dependent band gap evolutions are analyzed [Fig. S5 [42]]. Within uniaxial $\varepsilon_a$ and biaxial $\varepsilon_{ab}$ strains, the spin-down and total band gaps monotonically increase with increasing tensile strain and shrink under compression. In contrast, under uniaxial $\varepsilon_b$, the spin-down band gap is almost unchanged, while the total band gaps gradually decrease with tensile strain. Notably, under all strain conditions, the absolute values of opposite Mn magnetic moments are strictly equal [Fig. S6 [42]]. The spin-up band gaps are consistent with the total ones for all strain magnitudes. The AFM2 ground state and semiconducting fFIM character are preserved throughout, which is valuable for spintronic devices.

The MAE displays distinct strain-dependent behaviors [Figs. 5(d)-5(f)]. Under $\varepsilon_a$ strain [Fig. 5(d)], the energy differences $E_x$ - $E_z$ and $E_y$ - $E_z$ slightly decrease with increasing compressive strain, while tensile strain exhibits an opposite trend. For instance, at +3% tensile strain, $E_x$ - $E_z$ ($E_y$ - $E_z$) increases by 18 (81) μeV per unit cell. In contrast, when tensile $\varepsilon_b$ strain is applied, $E_x$ - $E_z$ decreases and that of $E_y$ - $E_z$ first increases and then decreases. Under -3% compressive strain, both $E_x$ - $E_z$ and $E_y$ - $E_z$ increase by 41 and 34 μeV per unit cell, respectively [Fig. 5(e)]. The $E_x$ - $E_z$ is almost insensitive to biaxial strain $\varepsilon_{ab}$, while $E_y$ - $E_z$ increases (decreases) significantly with tension (compression) [Fig. 5(f)]. Specifically, under 3% biaxial tensile strain, $E_y$ - $E_z$ increases by 67 μeV per unit cell. Comparing the MAE contributions with 3% biaxial strain and without strain [Figs. 3(g), S3 [42]], the positive (negative) contributions for $E_x$ - $E_z$ stem from the hybridized $d_{yz}$ and $d_{z^2}$, $d_{xy}$ and $d_{x^2-y^2}$, $d_{xy}$ and $d_{xz}$ ($d_{xz}$ and $d_{yz}$, $d_{yz}$ and $d_{x^2-y^2}$). For $E_y$ - $E_z$, the positive (negative) contributions stem from the hybridization between $d_{xy}$ and $d_{yz}$, $d_{xy}$ and $d_{x^2-y^2}$, $d_{xz}$ and $d_{x^2-y^2}$, $d_{xz}$ and $d_{z^2}$ ($d_{xz}$ and $d_{yz}$).

The relationship between spin splitting energy and applied strain is investigated via the energy magnitudes of the valence and conduction bands at the Y and Γ points [Figs. 5(g)-(i)]. The spin splitting of $Y_{VB}$ gradually decreases under increasing tensile strain, while $Y_{CB}$ and $\Gamma_{VB}$ show a continuous increase, with a maximum increment of 0.11 eV at +3% $\varepsilon_{ab}$. The $\Gamma_{CB}$ spin splitting decreases under $\varepsilon_a$ and $\varepsilon_{ab}$ but increases monotonically with tensile $\varepsilon_b$; for compressive -3%, its spin splitting is enhanced by 0.10 eV. Notably, throughout strained conditions, no imaginary frequencies are observed in phonon spectra [Fig. S7 [42]], indicating that the MnSe monolayer remains stable.

We further construct MnSe/$In_2Se_3$ heterostructures [Figs. 6(a) and 6(d)] to explore potential electrical control. The $In_2Se_3$ monolayer presents an indirect band structure with a gap value of 0.77 eV [Fig. S4(c) [42]]. In the upward ($P$↑) configuration of $In_2Se_3$/MnSe, the electronic states near the $E_F$ are primarily contributed by the MnSe layer, with clear energy separation between spin-up and spin-down channels [Figs. 6(b)-(c)]. The $In_2Se_3$-derived bands remain far from $E_F$ and barely interfere with the low-energy spin states of MnSe under the $P$↑ state, presenting an overall semiconducting feature for the heterostructure with spin-up, spin-down, and total band gaps of 0.40, 0.70, and 0.40 eV, respectively. When the ferroelectric polarization is switched to downward $P$↓ direction, substantial interactions occur between the $In_2Se_3$ and MnSe layers, and abundant hybridized electronic states cross the $E_F$, the heterostructure turns metallic. Meanwhile, the spin splitting of MnSe-originated bands is still preserved near $E_F$, guaranteeing persistent spin polarization. [Figs. 6(e)-(f))]. Note that the absolute values of the magnetic moments of the Mn atoms are strictly equal and the fFIM ground state is maintained under both $P$↑ and $P$↓ states.

The atom- and spin-resolved projected bands for MnSe and $In_2Se_3$ under $P$↑ and $P$↓ states are calculated [Figs. S8 and S9 [42]]. In the $P$↑ state, the spin-down Mn $d_{xy}$, spin-up $d_{xz}$ ($d_{z^2}$), spin-down Se $p_x$, spin-up Se $p_z$, from the MnSe layer, along with the spin-down Se $p_x$ from the $In_2Se_3$ layer, form prominent occupied bands near $E_F$, but no bands cross the $E_F$. When the polarization of $In_2Se_3$ in the heterostructure is switched to $P$↓, spin-down Mn $d_{xy}$, spin-down Se $p_x$ from the MnSe, and Se-$p_z$ from $In_2Se_3$ cross $E_F$, leading to the metallic state. To visualize the interfacial interaction and charge redistribution driven by ferroelectric polarization switching, we calculate the plane-averaged differential charge density ($\Delta\rho$) and the corresponding 3D differential charge density along the $z$ direction [Fig. S10 [42]]. The yellow and blue regions indicate electron accumulation and depletion, respectively. The $P$↓ state shows significantly stronger charge redistribution than the $P$↑ state at the interface. This enhanced charge redistribution alters the interfacial band alignment, which is responsible for the metallic state observed in the $P$↓ configuration.

### D. Magnetoresistance, electroresistance, and elastoresistance

In the light of the ferrimagnetic, ferroelectric, and ferroelastic orders alongside the substantial spin splitting, we constructed tunnel junctions to explore the triferroic transport performance [Fig. 1(b)]. The $In_2Se_3$/MnSe heterojunction is metallic or has a small band gap, making it suitable as electrodes, while MnSe exhibits a relatively broad band gap and can serve as a barrier. Four configurations were considered in the TFTJs: two $In_2Se_3$ ferroelectric polarizations up (D1) [Fig. 7(a)] or down (D2) [Fig. 7(b)], opposite (D3) [Fig. 7(c)], and transport along the perpendicular direction for comparison with D2 (D4) [Fig. 7(d)]. The transport coefficients in energy and momentum space are shown in Figs. 7(e)-7(h), 8(a)-8(d), and S11(a)-S11(b) [42]. The transport coefficients at the $E_F$, along with the spin filtering efficiency and the TMR, TER, and TAR ratios, are summarized in Tables 1-3.

The band structures of MnSe monolayer and the $In_2Se_3$/MnSe heterostructure [Fig. 3(a), 6(b), 6(c), 6(e), and 6(f)] indicate that both systems favor spin-up transport. High spin-filtering efficiencies ($\eta$ = 89%–100%) can be achieved in D1-$M_P$, D2-$M_P$, D2-$M_{AP}$, D3-$M_P$, and D4-$M_P$. This arises because the transmission in spin-up channel is stronger than that in spin-down channel for these configurations [Figs. 8(a)-8(d)], as verified by the projected local density of states (PLDOS): the spin-up channel exhibits a larger contribution through the extended region of the right electrode [Figs. 9(a)-9(d)]. Moderate efficiencies (72% for D1-$M_{AP}$ and -69% for D4-$M_{AP}$) and low efficiency (12% for D3-$M_{AP}$) stem from the modest and the small transmission difference between the two spin channels, respectively.

Furthermore, we examine triferroic manipulatable resistances of the constructed devices. The transmission when the magnetizations of the left and right parts of the device are parallel ($M_P$) is higher than that when they are antiparallel ($M_{AP}$) [Figs. 7(e)-7(h) and 8(a)-8(d)], as verified by that the electronic states through the extended region of the right electrode are fewer in $M_{AP}$ than $M_P$ [Figs. 9(a)-9(d)]. These yield TMRs of $2.98 \times 10^5$% for D1, $6.57 \times 10^2$% for D2, $2.28 \times 10^2$% for D3, and $1.07 \times 10^5$% for D4. Comparing the two configurations in D1, D2, and D3 can yield a TER of $2.93 \times 10^6$%-$6.97 \times 10^{14}$%. The ease of electron transport follows the order D2, D3, and D1 [Figs.

8(a)-8(d) and S11(a)-S11(b) [42]]. Comparing D4 with D2 gives the TAR. Along the transport direction of D4, the density of electronic states is more than that of D2 [Figs. 9(b) and 9(d)], resulting in TARs of $7.95 \times 10^4\%$ for D4-$M_P$ vs D2-$M_P$ and $4.63 \times 10^2\%$ for D4-$M_{AP}$ vs D2-$M_{AP}$. These results provide a route to coexisting magnetic-, electric-, and elastic-tunable resistance states.

## IV. CONCLUSION

In summary, we demonstrate that hidden-phase wurtzite MnSe monolayer is an intrinsic fFIM, arising from two inequivalent Mn sublattices with opposite magnetic moments that are not connected by any symmetry operation. The monolayer possesses favorable thermal, dynamical, and mechanical stability as well as high mechanical flexibility. It is a UMS with spin-up, spin-down, and total band gaps of 1.82, 2.20, and 1.82 eV, respectively, and exhibits substantial spin splitting energies of 0.49 eV at the VBM (Y point) and 0.20 eV at the CBM (Γ point). The system exhibits perpendicular magnetic anisotropy (528.60 μeV per unit cell). This magnetic order coexists with intrinsic ferroelectricity and ferroelasticity with moderate barriers of $7.6 \times 10^{-3}$ and 0.10 eV/f.u., respectively, rendering the monolayer a rare single-phase triferroic material. The in-plane ferroelectric polarization reaches $4.63 \times 10^{-10}$ C/m and the ferroelastic signal intensity is 61%. The ground fFIM UMS features are robust against uniaxial and biaxial strains, while the splitting energy can be modulated. With $In_2Se_3$ monolayer, we achieve nonvolatile electrical control of a semiconductor-to-metal transition via ferroelectric switching. The constructed tunnel junctions display TMR of $2.98 \times 10^5\%$, TER of $6.97 \times 10^{14}\%$, TAR of $7.95 \times 10^4\%$, and nearly perfect spin filtering efficiency (~100%). Our results establish an illuminating route for multifield-controllable spintronics, electronics, and straintronics.

## ACKNOWLEDGMENTS

This work was supported by the National Natural Science Foundation of China (Grant No. 12374002), the Natural Science Foundation of Henan Province (No. 242300420996 and No. 262300422670), and the project Quantum materials for applications in sustainable technologies (QM4ST), funded as project No. CZ.02.01.01/00/22_008/0004572 by Programme Johannes Amos Commenius, call Excellent Research. The authors acknowledge Beijing PARATERA Technology Co., LTD for providing high-performance and AI computing resources for contributing to the research results reported within this paper. URL: http://www.cloud.paratera.com.

## DATA AVAILABILITY

The data that support the findings of this article are not publicly available upon publication because it is not technically feasible and/or the cost of preparing, depositing, and hosting the data would be prohibitive within the terms of this research project. The data are available from the authors upon reasonable request.

## Figures and Tables

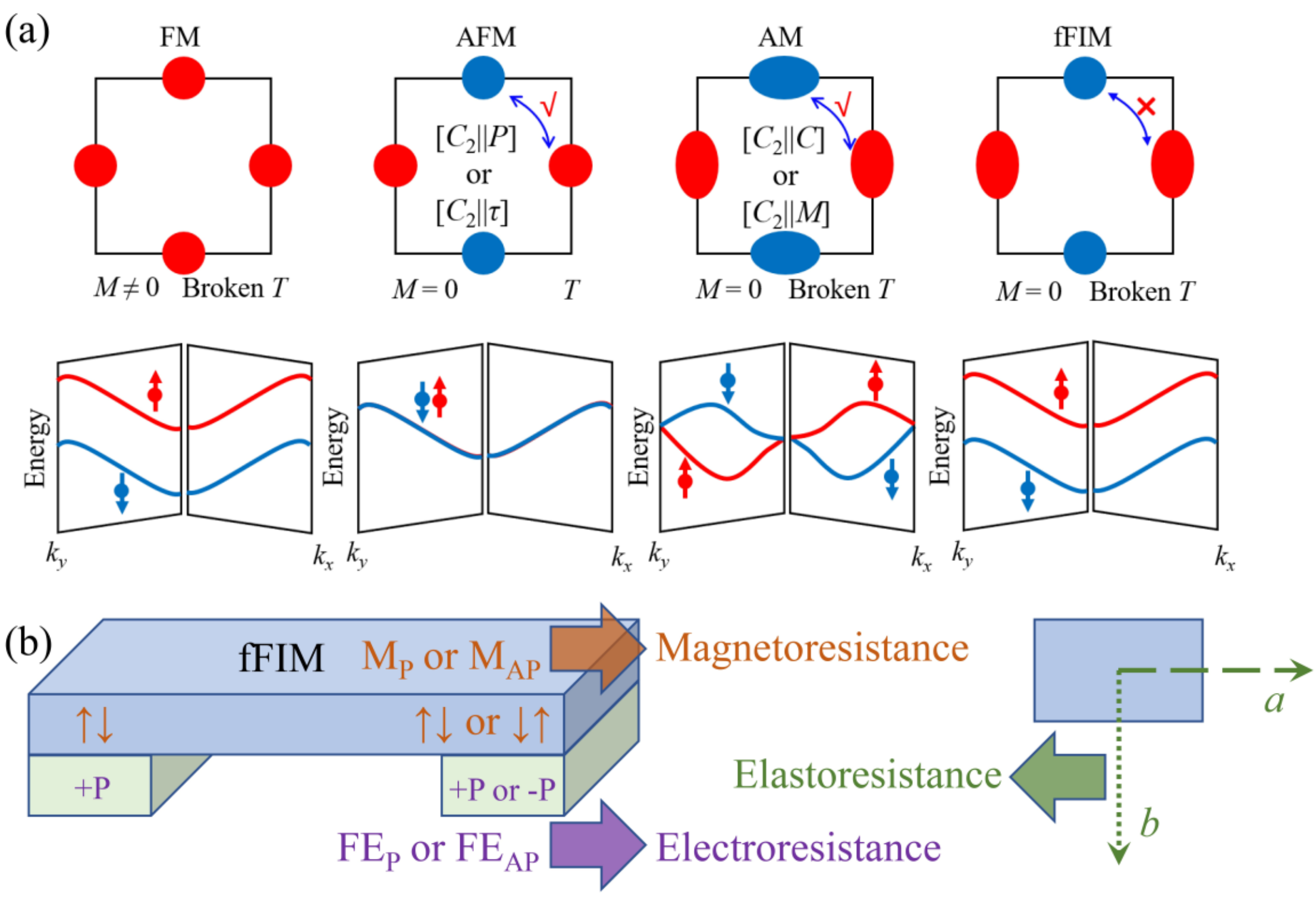

FIG. 1. (a) Four categories of collinear magnetic order and corresponding band splitting. Ferromagnets (FMs) break time-reversal symmetry ($T$) and exhibit non-zero macroscopic magnetization ($M \neq 0$) with spin-splitting bands. Antiferromagnets (AFMs) preserve $T$ and exhibit zero macroscopic magnetization ($M = 0$) with spin-degenerate bands, where the magnetic sublattices are connected by inversion ($P$) and translation ($\tau$). Altermagnets (AMs) break $T$ with zero net magnetization ($M = 0$) and exhibit anisotropic spin-splitting bands, where the magnetic sublattices are connected by rotation ($C$) or mirror ($M$). Fully compensated ferrimagnets (fFIMs) break $T$ with zero net magnetization ($M = 0$) and exhibit ferromagnetic-like spin-splitting bands, where the magnetic sublattices have no symmetry connections. (b) Schematic illustration of the fFIM multistate transport. $M_P$ and $M_{AP}$ indicate the parallel and antiparallel magnetization alignments, respectively. +$P$ and -$P$ indicate the upward and downward ferroelectric polarization directions of the ferroelectric layer. $a$ and $b$ correspond to transport along the perpendicular directions, which denotes the initial and final ferroelastic states, respectively. These configurations enable the realization of magnetoresistance, electroresistance, and elastoresistance effects.

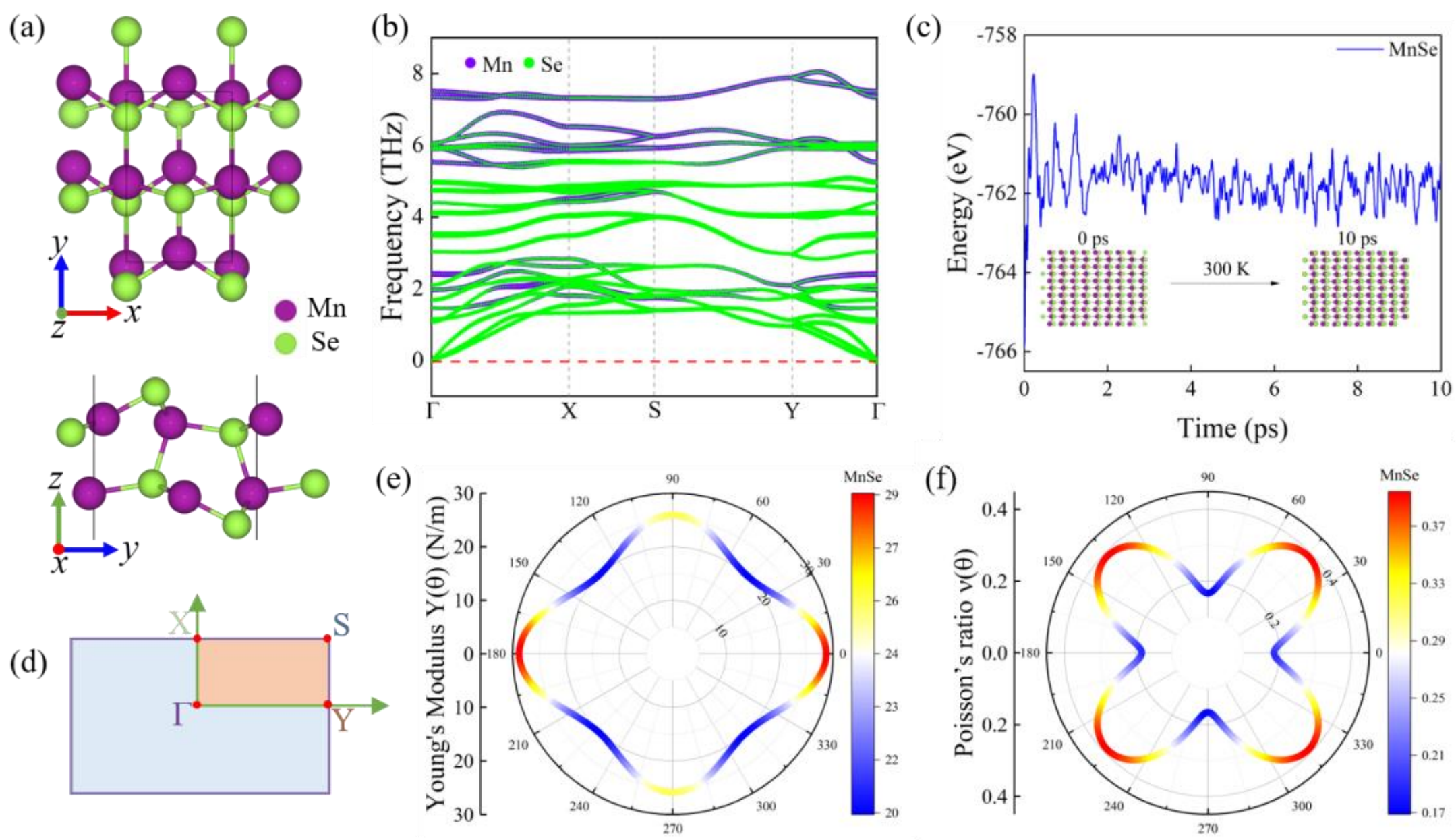

FIG. 2. (a) Top and side views of monolayer MnSe. (b) Projected phonon dispersion spectra and (c) AIMD results. (d) First Brillouin zone of monolayer MnSe. (e) In-plane Young's modulus and (f) Poisson's ratio for monolayer MnSe as a function of angle $\theta$, where $\theta = 0$ corresponds to the $x$ direction.

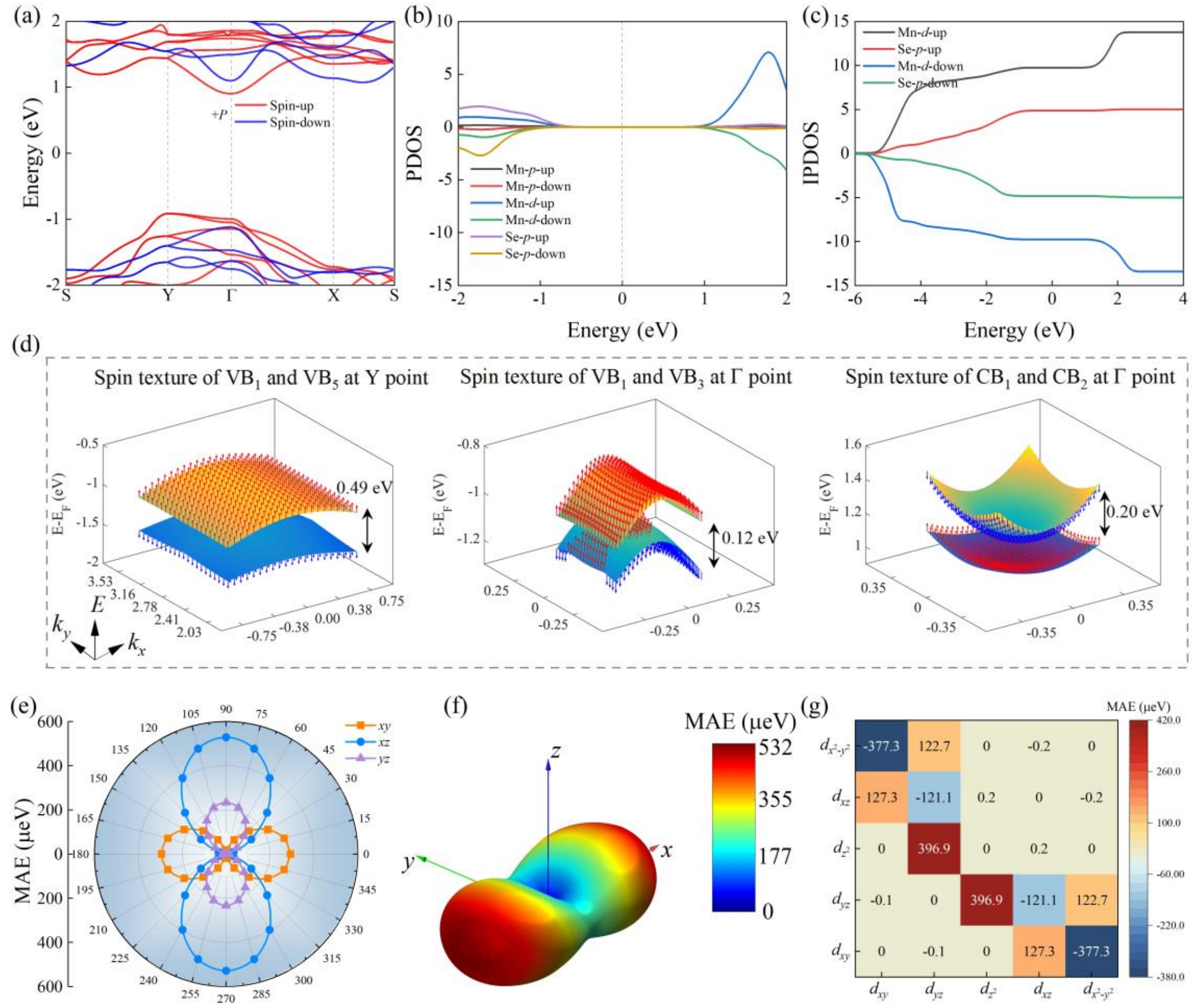

FIG. 3. (a) Band structures of monolayer MnSe. (b) The spin-resolved projected density of states (PDOS) and (c) the integrated projected density of states (IPDOS) of MnSe. (d) Spin textures of the first and fifth valence bands ($VB_1$ and $VB_5$) at the Y point, the first and third valence bands ($VB_1$ and $VB_3$) at the Γ point, and first and second conduction bands ($CB_1$ and $CB_2$) in the Γ point. The magnetic moments and spin polarization are oriented along the *z*-axis. (e) The angular dependence of magnetic anisotropy energy (MAE) for MnSe in the *xy*, *xz*, and *yz* planes, showing the in-plane and out-of-plane variations. (f) 3D visualization of the MAE angular dependence in the whole space. (g) Mn-3*d*-orbital-resolved MAE for $E_x$ - $E_z$.

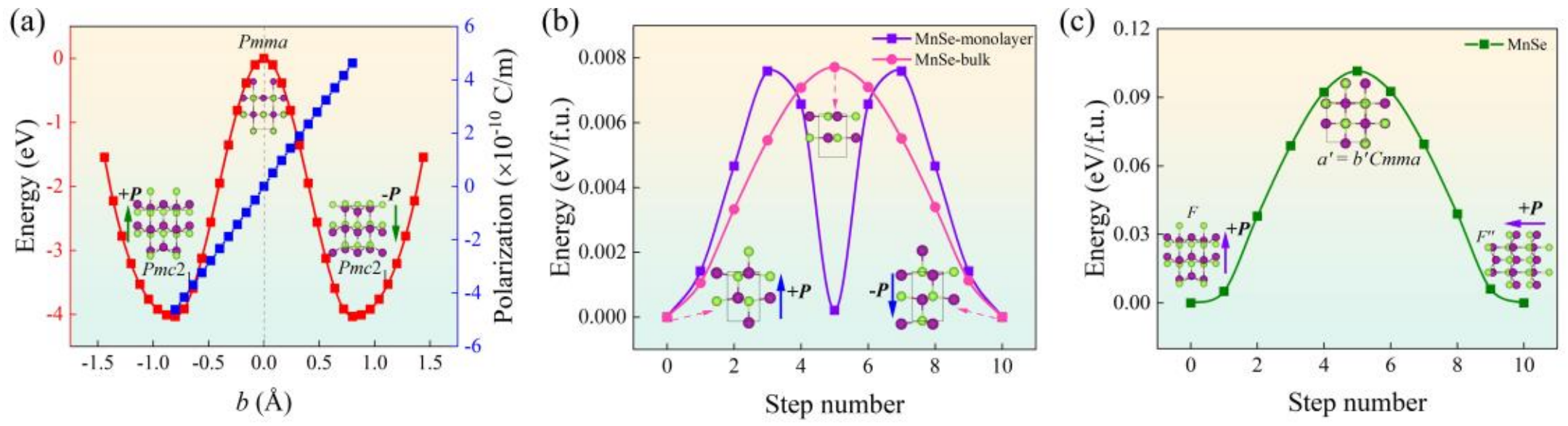

FIG. 4. (a) Relative energy (red line) and polarization value (blue line) as a function of the displacement amplitude $b$ of monolayer MnSe, where $+P$ and $-P$ denote the opposite ferroelectric states. (b) Pathway and barriers for monolayer and bulk MnSe, where the arrows denote the initial, transition, and final states of bulk MnSe. (c) Energy barriers of ferroelastic switching as a function of step number for monolayer MnSe.

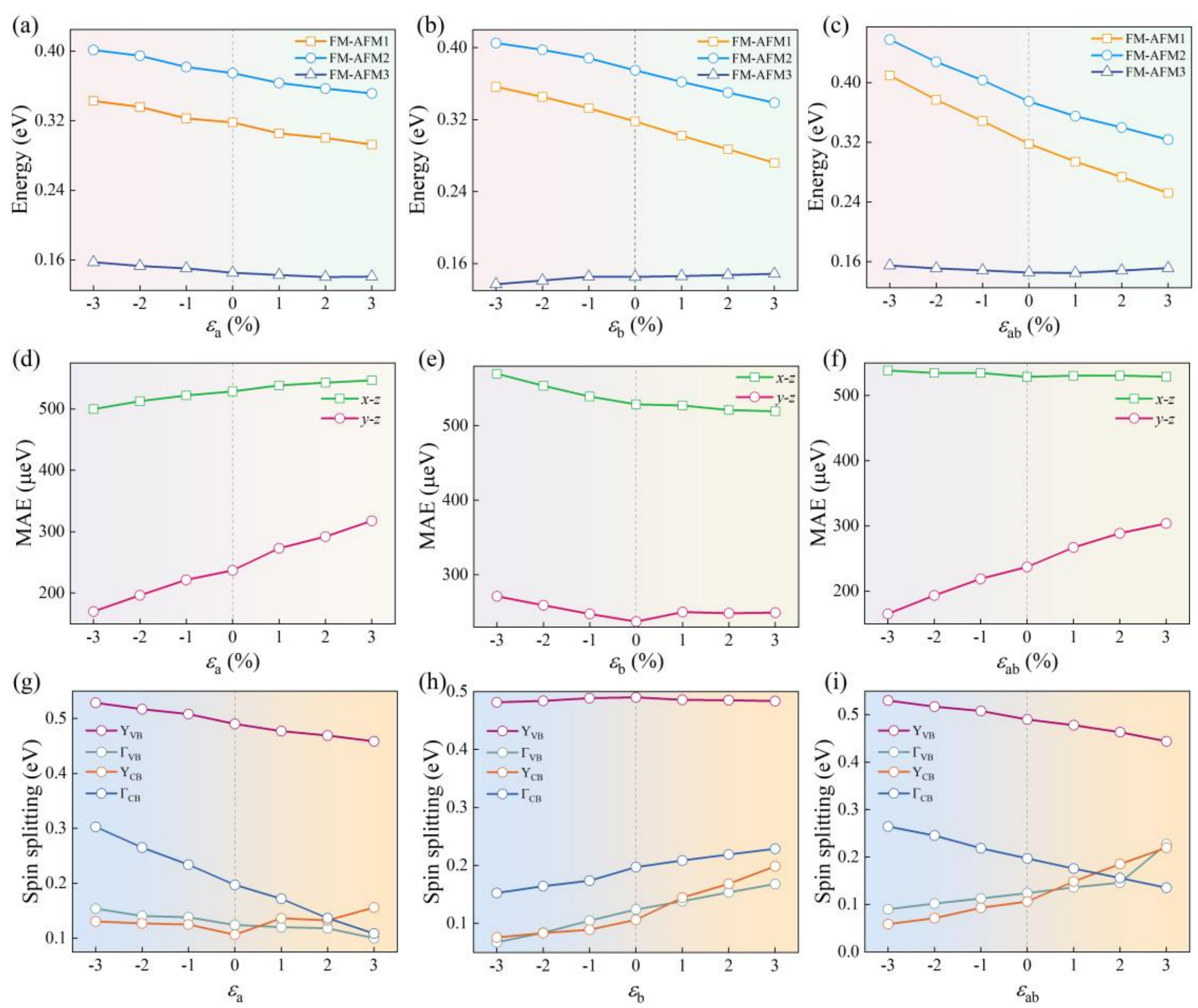

FIG. 5. (a)-(c) The energy difference ($\Delta E = E_{FM} - E_{AFM}$), (d)-(f) the total MAE, and (g-i) the spin splitting of the valence and conduction bands at the Y and Γ points as a function of strain for the monolayer MnSe unit cell under strains.

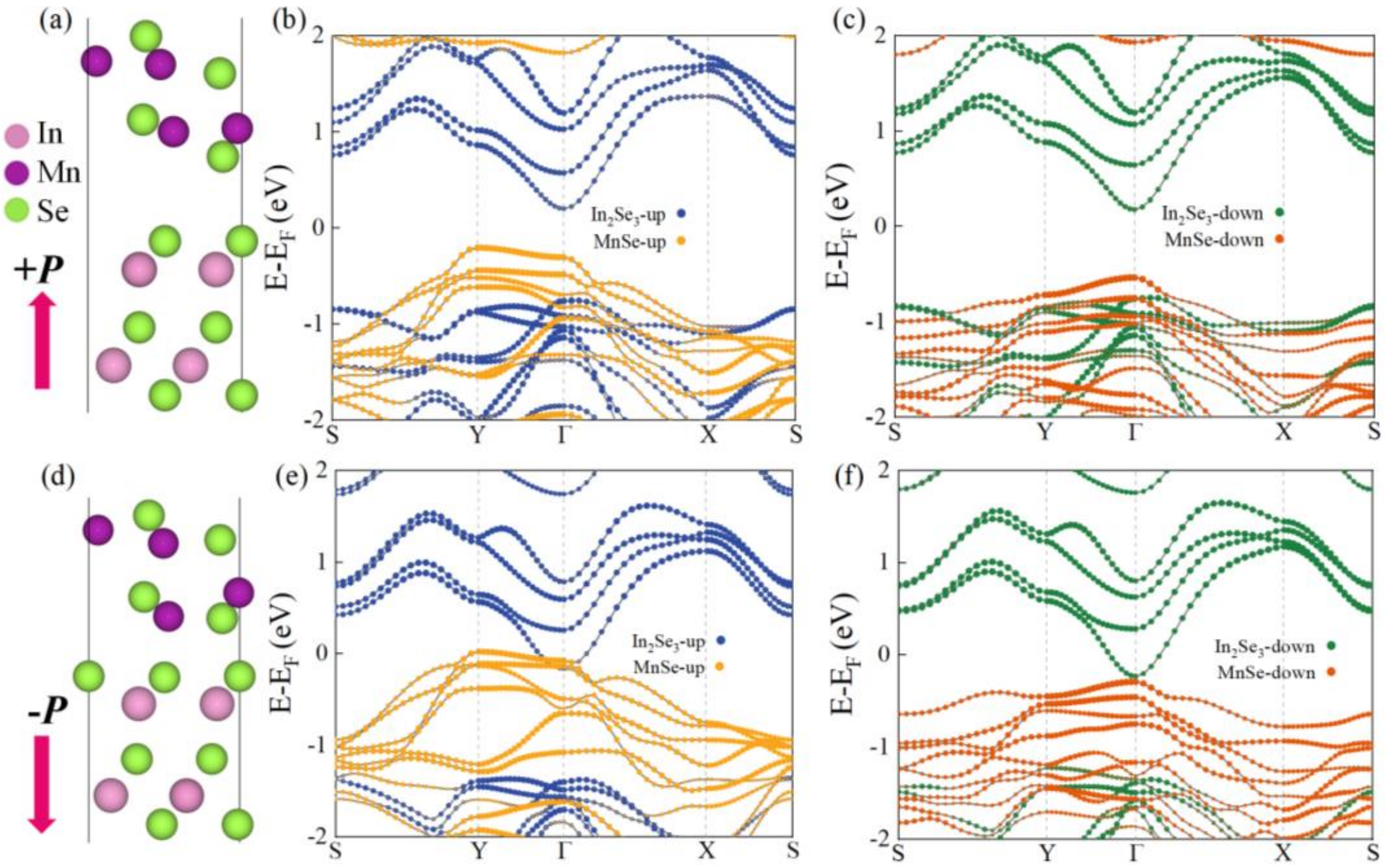

FIG. 6. Crystal structures and projected band structures in spin-up and spin-down channels with (a-c) upward (+*P*) and (d-f) downward (-*P*) ferroelectric polarization directions for the $In_2Se_3$/MnSe heterostructure. The blue/green ($In_2Se_3$) and yellow/orange (MnSe) dots denote layer-resolved contributions. The Fermi level is set to 0.

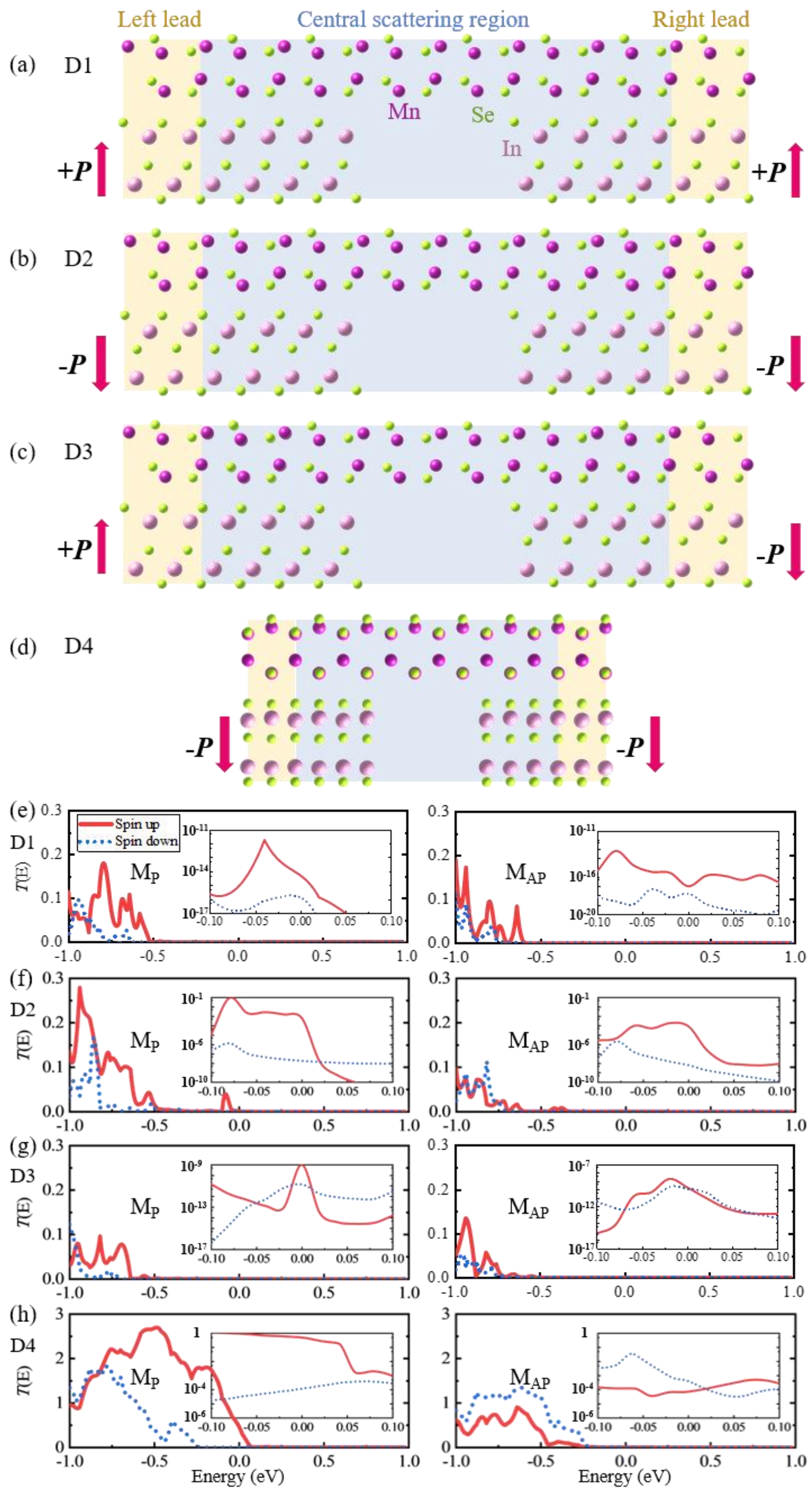

FIG. 7. (a-d) Structural models of the $MnSe/In_2Se_3$ TFTJs. (e-h) Transmission coefficients as a function of energy in parallel ($M_P$) and antiparallel ($M_{AP}$) magnetization configurations.

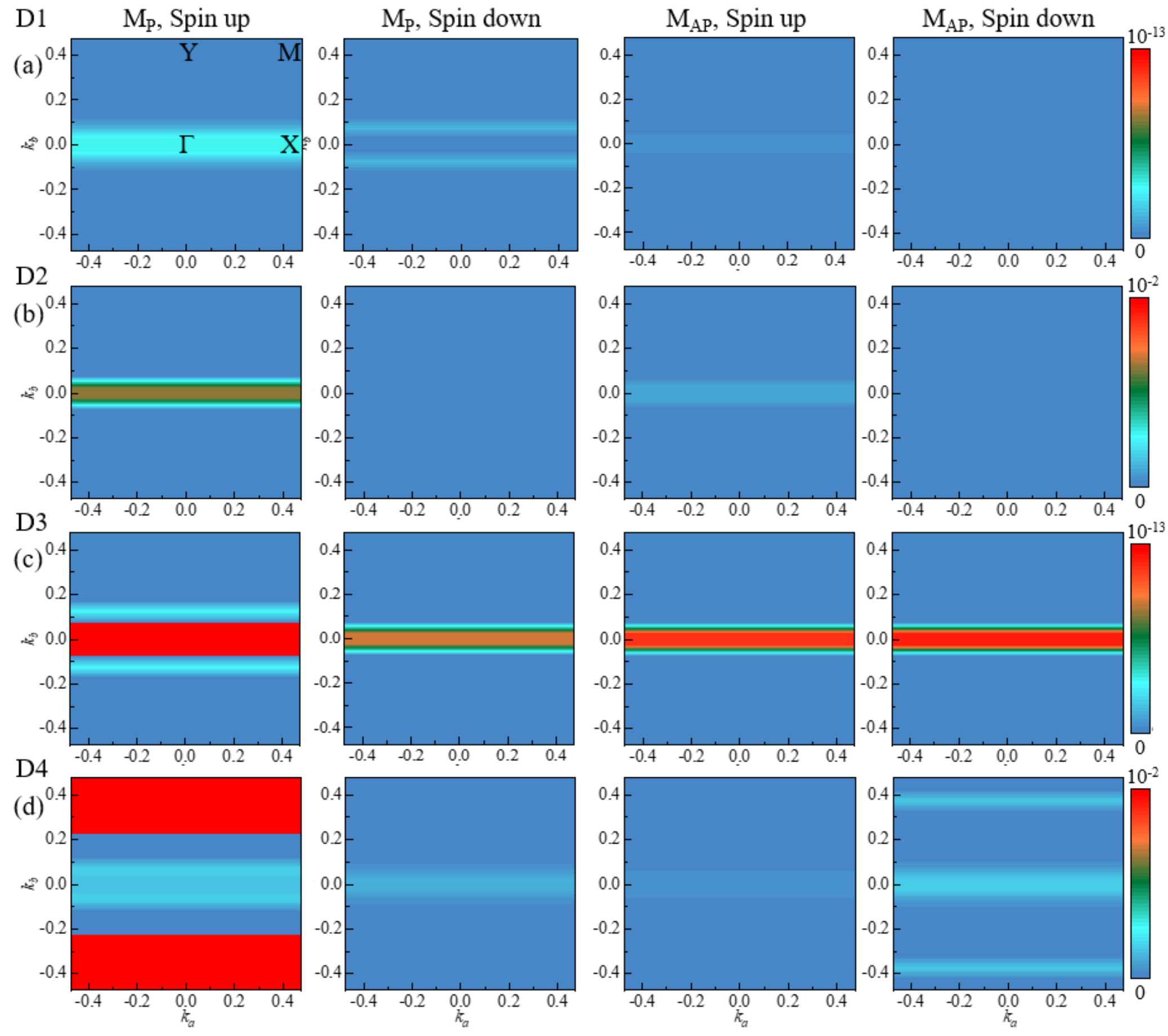

FIG. 8. (a-d) The momentum-resolved transmission spectrum in the 2D Brillouin zone across MnSe/$In_2Se_3$ TFTJs.

Table 1. Spin-dependent electron transmission $T$, spin filtering efficiency $\eta$, and TMR ratios across MnSe/$In_2Se_3$ TFTJs.

| | $M_P$ | | | | $M_{AP}$ | | | | TMR (%) |
|---|---|---|---|---|---|---|---|---|---|
| | $T_\uparrow$ | $T_\downarrow$ | $T$ | $\eta$ (%) | $T_\uparrow$ | $T_\downarrow$ | $T$ | $\eta$ (%) | |
| D1 | $1.95 \times 10^{-14}$ | $3.35 \times 10^{-15}$ | $3.55 \times 10^{-14}$ | 89 | $1.02 \times 10^{-17}$ | $1.65 \times 10^{-18}$ | $1.19 \times 10^{-17}$ | 72 | $2.98 \times 10^5$ |
| D2 | $6.28 \times 10^{-4}$ | $1.84 \times 10^{-8}$ | $6.28 \times 10^{-4}$ | 100 | $8.30 \times 10^{-5}$ | $6.36 \times 10^{-9}$ | $8.30 \times 10^{-5}$ | 100 | $6.57 \times 10^2$ |
| D3 | $1.03 \times 10^{-9}$ | $1.41 \times 10^{-11}$ | $1.04 \times 10^{-9}$ | 98 | $1.77 \times 10^{-10}$ | $1.40 \times 10^{-10}$ | $3.17 \times 10^{-10}$ | 12 | $2.28 \times 10^2$ |
| D4 | 0.50 | $1.17 \times 10^{-4}$ | 0.50 | 100 | $7.12 \times 10^{-5}$ | $3.95 \times 10^{-4}$ | $4.67 \times 10^{-4}$ | -69 | $1.07 \times 10^5$ |

Table 2. TER ratios with various ferroelectric configurations across $MnSe/In_2Se_3$ TFTJs.

| | ($D2\text{-}M_P$, $D1\text{-}M_P$) | ($D2\text{-}M_{AP}$, $D1\text{-}M_{AP}$) | ($D3\text{-}M_P$, $D1\text{-}M_P$) | ($D3\text{-}M_{AP}$, $D1\text{-}M_{AP}$) | ($D2\text{-}M_P$, $D3\text{-}M_P$) | ($D2\text{-}M_{AP}$, $D3\text{-}M_{AP}$) |
|---|---|---|---|---|---|---|
| TER (%) | $1.77 \times 10^{12}$ | $6.97 \times 10^{14}$ | $2.93 \times 10^{6}$ | $2.66 \times 10^{9}$ | $6.04 \times 10^{7}$ | $2.62 \times 10^{7}$ |

Table 3. TAR ratios with various ferroelastic configurations across $MnSe/In_2Se_3$ TFTJs.

| | ($D4\text{-}M_P$, $D2\text{-}M_P$) | ($D4\text{-}M_{AP}$, $D2\text{-}M_{AP}$) |
|---|---|---|
| TAR (%) | $7.95 \times 10^{4}$ | $4.63 \times 10^{2}$ |

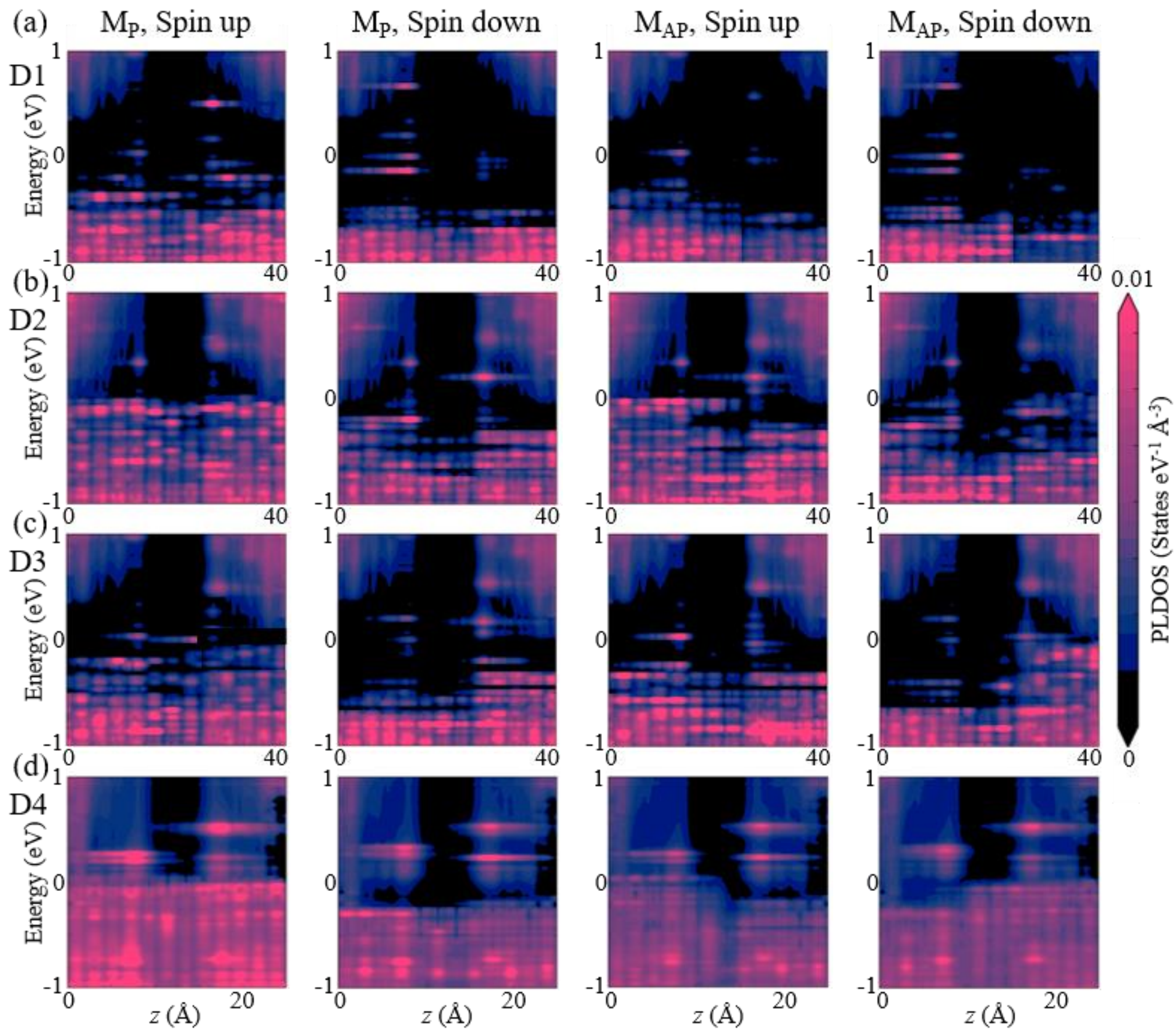

FIG. 9. The spin-resolved projected local density of states (PLDOS) across $MnSe/In_2Se_3$ TFTJs. The electronic contributions stem from the left electrodes serving as the source.